\documentclass[sigconf, dvipsnames, natbib=true]{acmart}
\makeatletter \gdef\@ACM@checkaffil{} \makeatother

\usepackage{geometry}
\usepackage{graphicx}
\usepackage{pgfplots}
\pgfplotsset{compat=1.17}
\usepackage{tikz-cd}
\usepackage{tikz}
\usepackage{hyperref}
\usepackage{lipsum}
\usepackage{listings}
\usepackage{subcaption}
\usepackage{geometry}
\usepackage{enumitem}
\usepackage{xcolor}
\usepackage{caption}
\usepackage{tcolorbox, xcolor, soul, xspace}
\usepackage{titlesec}

\titlespacing*{\section}{0pt}{0.1\baselineskip}{0.2\baselineskip}

\usepackage{tcolorbox}
\tcbuselibrary{listings, breakable, skins, theorems}
\usepackage{soul} 

\sethlcolor{yellow!30}

\usepackage{booktabs}
\usepackage{colortbl}
\usepackage{xcolor}

\rowcolors{2}{gray!5}{white}

\usepackage{pgfplotstable}
\usepgfplotslibrary{colorbrewer}
\pgfplotsset{compat=1.17}

\usepgfplotslibrary{groupplots}
\let\origthelstnumber\thelstnumber
\makeatletter
\newcommand*\Suppressnumber{%
  \lst@AddToHook{OnNewLine}{%
    \let\thelstnumber\relax%
     \advance\c@lstnumber-\@ne\relax%
    }%
}

\newcommand*\Reactivatenumber[1]{%
  \lst@AddToHook{OnNewLine}{%
   \let\thelstnumber\origthelstnumber%
   \setcounter{lstnumber}{\numexpr#1-1\relax}%
   }%
}

\pgfplotsset{compat=1.17}
 \usepackage{etoolbox}
\makeatletter
\patchcmd{\@verbatim}
  {\verbatim@font}
  {\verbatim@font\small}
  {}{}
\makeatother

\hypersetup{
   breaklinks=true,   
   colorlinks=true,   
}

\AtBeginDocument{%
  \providecommand\BibTeX{{%
    \normalfont B\kern-0.5em{\scshape i\kern-0.25em b}\kern-0.8em\TeX}}}


\copyrightyear{2025}
\acmYear{2025}
\setcopyright{rightsretained}
\acmConference[arXiv '25]{arXiv Submission}{Apr 22, 2025}{Virginia Tech, USA}
\acmBooktitle{arXiv '25, Apr 22, 2025, Virginia Tech, VA, USA}

\def\BibTeX{{\rm B\kern-.05em{\sc i\kern-.025em b}\kern-.08em
    T\kern-.1667em\lower.7ex\hbox{E}\kern-.125emX}}

\renewcommand\footnotetextcopyrightpermission[1]{} 
\newcommand{\CB}{compute-bound\xspace}
\newcommand{\BB}{bandwidth-bound\xspace}

\newcommand{\myGPU}{NVIDIA RTX 3080\xspace}
\newcommand{\gptFOURoMini}{\texttt{gpt-4o-mini}\xspace}

\newcommand{\gptOone}{\texttt{o1}\xspace}
\newcommand{\gptFOURFIVE}{\texttt{gpt-4.5-preview}\xspace}

\newcommand{\gptOthreeMiniHigh}{\texttt{o3-mini-high}\xspace}

\newcommand{\classificationQuestion}{RQ1\xspace}
\newcommand{\zeroShotQuestion}{RQ2\xspace}
\newcommand{\fewShotQuestion}{RQ3\xspace}
\newcommand{\fineTuningQuestion}{RQ4\xspace}

\hypersetup{
    breaklinks=true,
    colorlinks=true,
    citecolor=MidnightBlue,    
    linkcolor=BrickRed,        
    urlcolor=Blue,             
}

\begin{document}

\title{Can Large Language Models Predict Parallel Code Performance?}

\author{Gregory Bolet}
\email{gbolet@vt.edu}
\affiliation{Virginia Tech, USA}

\author{Giorgis Georgakoudis}
\email{georgakoudis1@llnl.gov}
\affiliation{LLNL, USA}

\author{Harshitha Menon}
\email{harshitha@llnl.gov}
\affiliation{LLNL, USA}

\author{Konstantinos Parasyris}
\email{parasyris1@llnl.gov}
\affiliation{LLNL, USA}

\author{Niranjan Hasabnis}
\email{niranjan@codemetal.ai}
\affiliation{Code Metal, USA}

\author{Hayden Estes}
\email{haydenve@vt.edu}
\affiliation{Virginia Tech, USA}

\author{Kirk W. Cameron}
\email{cameron@vt.edu}
\affiliation{Virginia Tech, USA}

\author{Gal Oren}
\email{galoren@stanford.edu}
\affiliation{Stanford University, Technion, USA}

\renewcommand{\shortauthors}{Bolet et al.}

\begin{abstract}
Accurate determination of the performance of parallel GPU code typically requires execution-time profiling on target hardware -- an increasingly prohibitive step due to limited access to high-end GPUs.
This paper explores whether Large Language Models (LLMs) can offer an alternative approach for GPU performance prediction without relying on hardware.
We frame the problem as a \textit{roofline classification} task: given the source code of a GPU kernel and the hardware specifications of a target GPU, can an LLM predict whether the GPU kernel is compute-bound or bandwidth-bound?

For this study, we build a balanced dataset of 340 GPU kernels, obtained from HeCBench benchmark and written in CUDA and OpenMP, along with their ground-truth labels obtained via empirical GPU profiling.
We evaluate LLMs across four scenarios: (1) with access to profiling data of the kernel source, (2) zero-shot with source code only, (3) few-shot with code and label pairs, and (4) fine-tuned on a small custom dataset.
Our results show that state-of-the-art LLMs have a strong understanding of the Roofline model, achieving 100\% classification accuracy when provided with explicit profiling data.
We also find that reasoning-capable LLMs significantly outperform standard LLMs in zero- and few-shot settings, achieving up to 64\% classification accuracy of GPU source codes, without any profiling information.
Lastly, we find that model accuracy does not benefit meaningfully from few-shot prompting compared to zero-shot, and that LLM fine-tuning will require much more data than what we currently have available.
This work is among the first to use LLMs for source-level roofline performance prediction via classification, and illustrates their potential to guide optimization efforts when runtime profiling is infeasible.
Our findings suggest that with better datasets and prompt strategies, LLMs could become practical tools for HPC performance analysis and performance portability.
Code and datasets are publicly available at \url{https://github.com/Scientific-Computing-Lab/ParallelCodeEstimation}.
\end{abstract}

\keywords{Roofline model, LLMs, CUDA, OpenMP, GPU, Performance\vspace{-0.3cm}}


\maketitle


\section{Introduction}

High-Performance Computing (HPC) applications increasingly rely on GPUs for acceleration, yet ensuring \emph{performance portability} across diverse GPU architectures has become a pressing challenge~\cite{rooflineERT1,kokkos3,raja}. 
Modern supercomputers feature a mix of GPUs from different vendors, and codes optimized for one platform may not achieve optimal performance on another. 
Identifying performance bottlenecks, whether a kernel is limited by computation or by memory throughput, is crucial to guide optimizations. 
Traditionally, this determination requires running detailed profiles on target hardware~\cite{laksonoadhiantoHPCTOOLKITToolsPerformance2009,josemorgadoCARMToolCacheAware2024,lechenLandscapeChallengesHPC2024}, but access to cutting-edge GPUs is often limited and expensive~\cite{Cohan2024,seonholeeForecastingGPUPerformance2024}. 
There is a growing need for \emph{hardware-free performance prediction} methods that can provide insights without extensive benchmarking on physical devices.
This work addresses that need by exploring a novel, static approach to classify GPU kernel performance.

A well-established framework for reasoning about code performance is the \emph{Roofline model}~\cite{rooflineModel}. 
The Roofline model correlates a kernel’s \emph{arithmetic intensity (\textbf{AI})} (operations per byte of memory traffic) with hardware peak performance (operations per second) to determine a performance ceiling. 
Kernels with low AI tend to be \emph{Bandwidth-Bound (\textbf{BB})} (limited by memory bandwidth), while those with high AI are \emph{Compute-Bound (\textbf{CB})} (limited by the processor’s peak FLOPs)~\cite{rooflineERT2}. 
By plotting a kernel’s AI against achievable performance, one can visualize whether performance is capped by bandwidth or compute limits. 
This understanding of AI and performance relative to a target hardware roofline can then be used to intuitively guide kernel optimization decisions to reach peak system performance~\cite{yujungloRooflineModelToolkit2014,rooflineERT1}.
However, obtaining a kernel’s AI and achieved performance usually entails \emph{profiling} -- measuring runtime operations and memory transfers on the actual GPU. 
Similarly, previous GPU performance modeling works also depend on performance counter metrics ~\cite{lorenzbraunSimpleModelPortable2020,arunavodeyRelativePerformancePrediction2024}. 
This requirement hampers rapid performance analysis, especially when developers lack access to the target hardware. 
Static analysis works ~\cite{gargialavaniPredictingExecutionTime2018,guerreiroGPUStaticModeling2019,fanPredictableGPUsFrequency2019,alavaniApproachEstimatePower2020} have shown potential for GPU power and execution time prediction, however, these still require hardware access for micro-benchmarking.

\emph{Large Language Models (LLMs)} offer a fresh opportunity to address this problem of requiring hardware access and runtime information to predict performance. 
Recent advances in code-focused LLMs have demonstrated remarkable capabilities in understanding and generating code. 
These models can perform tasks such as summarization, translation, bug fixing, and optimization~\cite{chen2021codex, xu2022systematic,binleiCreatingDatasetHighPerformance2023}. 
In HPC contexts, early studies have begun applying LLMs to assist in automatic parallelization~\cite{wang2023llmomp}, GPU kernel execution time prediction~\cite{minh-khoinguyen-nhatLLMPerfGPUPerformance2024}, performance slowdown prediction~\cite{danielnicholsModelingParallelPrograms2023}, and performance optimization~\cite{bowencuiLargeLanguageModels2025,danielnicholsPerformanceAlignedLLMsGenerating2024}.
These developments suggest that LLMs not only understand code syntax but may also capture deeper semantic features and performance-affecting structures.
The advantage of the Roofline model is that it can intuitively guide \textit{developers} -- and thus, LLMs, towards an optimal program.
Our work is the beginning of an effort that seeks to leverage this nature of the Roofline model to guide LLMs towards delivering performant code.
However, integration of \emph{modern LLMs} into Roofline-based performance analysis remains unexplored. 
This work seeks to establish a base understanding of how well LLMs recall and use the Roofline model to predict program performance.

We build upon the trends of: Roofline modeling, static prediction, and LLM-driven code understanding -- by framing GPU performance prediction as a binary classification task. 
We set our sights on a simple sub-problem within Roofline modeling: classifying a program as BB or CB based on its AI.
We first assess how well LLMs can classify kernels when given known AI values; expecting the LLMs to do simple calculations and reasoning to arrive at an answer.
We then profile CUDA-based and OpenMP-based GPU kernels from the HeCBench benchmark suite~\cite{hecbenchGithub,hecbenchPaper} and obtain their ground-truth AI classification labels.
Given a GPU kernel, we have LLMs predict whether it is BB or CB using only the source code and hardware/execution specs.
By comparing LLM predictions to our ground-truth labels, we assess whether LLMs can intuit about AI, and if they can serve as practical tools for source-code-level performance classification.
From this point forward, we refer to this task of classifying the AI of a program (to be BB or CB) as the \emph{roofline classification task}.

\noindent\textbf{Research Questions.}
\label{sec:RQs}
To study how LLMs can be applied to the roofline classification task, we ask the following questions:
\begin{itemize}[leftmargin=*,topsep=0pt]
    \item \textbf{\classificationQuestion (Baseline Roofline Classification):} Can LLMs classify codes \emph{well} when given the hardware roofline, and arithmetic intensity values?
    \item \textbf{\zeroShotQuestion (Zero-Shot Classification):} Can LLMs classify codes \emph{well} when \emph{given their source code, hardware specs, and minimal instructions}?
    \item \textbf{\fewShotQuestion (Few-Shot Classification):} Can LLMs classify codes \emph{well} when given their source code, hardware specs, and a few examples of codes with their expected classifications?
    \item \textbf{\fineTuningQuestion (Fine-Tuned Classification):} Can we fine-tune LLMs for roofline classification? 
\end{itemize}

\noindent\textbf{Contributions.}
The contributions of this paper are:
\begin{enumerate}[leftmargin=*,topsep=0pt]
    \item To the best of our knowledge, this is the first systematic evaluation of LLMs for the GPU roofline classification task.
    \item We demonstrate that existing state-of-the-art (SoTA) LLMs are highly accurate in roofline classification when presented with the profiled measurements of a kernel.
    \item We find that small, reasoning-capable LLMs can significantly outperform their larger counterparts as well as traditional non-reasoning LLMs in roofline classification.
    \item Our experiments suggest that there exists room for improvement of current LLMs in roofline classification.
\end{enumerate}


\section{Our Approach}

To gauge the efficacy of LLMs for the task of roofline classification, we empirically measure the AI of existing programs to form a ground-truth database. 
We query LLMs, asking for roofline classifications based on a program's source code and target hardware specs.
We focus on GPU codes for two reasons: 1) because GPU workloads have become ubiquitous, thus this work can have a broad impact, and 2) because (in contrast to CPU-based codes) their source codes do not need to be manually instrumented to record runtime performance metrics for the roofline model.
In the following, we describe the process of profiling, data collection, and pre-processing for evaluation.

\subsection{Sampled Codes \& Hardware}
We built and profiled 446 CUDA and 303 OpenMP offload GPU codes from the HeCBench suite \cite{hecbenchGithub,hecbenchPaper}.
The target hardware for profiling was an \myGPU with 10GB of RAM; selected due to physical access to the hardware.
We profiled the GPU \textit{kernels} within these HeCBench codes and were able to use their empirical measurements to form ground-truth AI classification labels.
We capture single-precision (SP-FLOP) float, double-precision (DP-FLOP) float, and integer operations (INTOP) counts alongside execution time and number of global memory read/write operations for all the kernels within each program.
To save time on profiling overhead of multiple/repeated kernel invocations, we only profile the first execution of each kernel within a program.
From these metrics, we classify each of the kernels as BB or CB, relative to the three arithmetic operation rooflines: SP-FLOP, DP-FLOP, or INTOP corresponding to the major types of operations performed by the kernels.
If a kernel is BB in all 3 arithmetic operations, we consider it BB for classification; otherwise if there exists at least 1 operation type where the kernel is CB, we consider it CB.

 \begin{figure}
     \centering
     \includegraphics[width=1.0\linewidth]{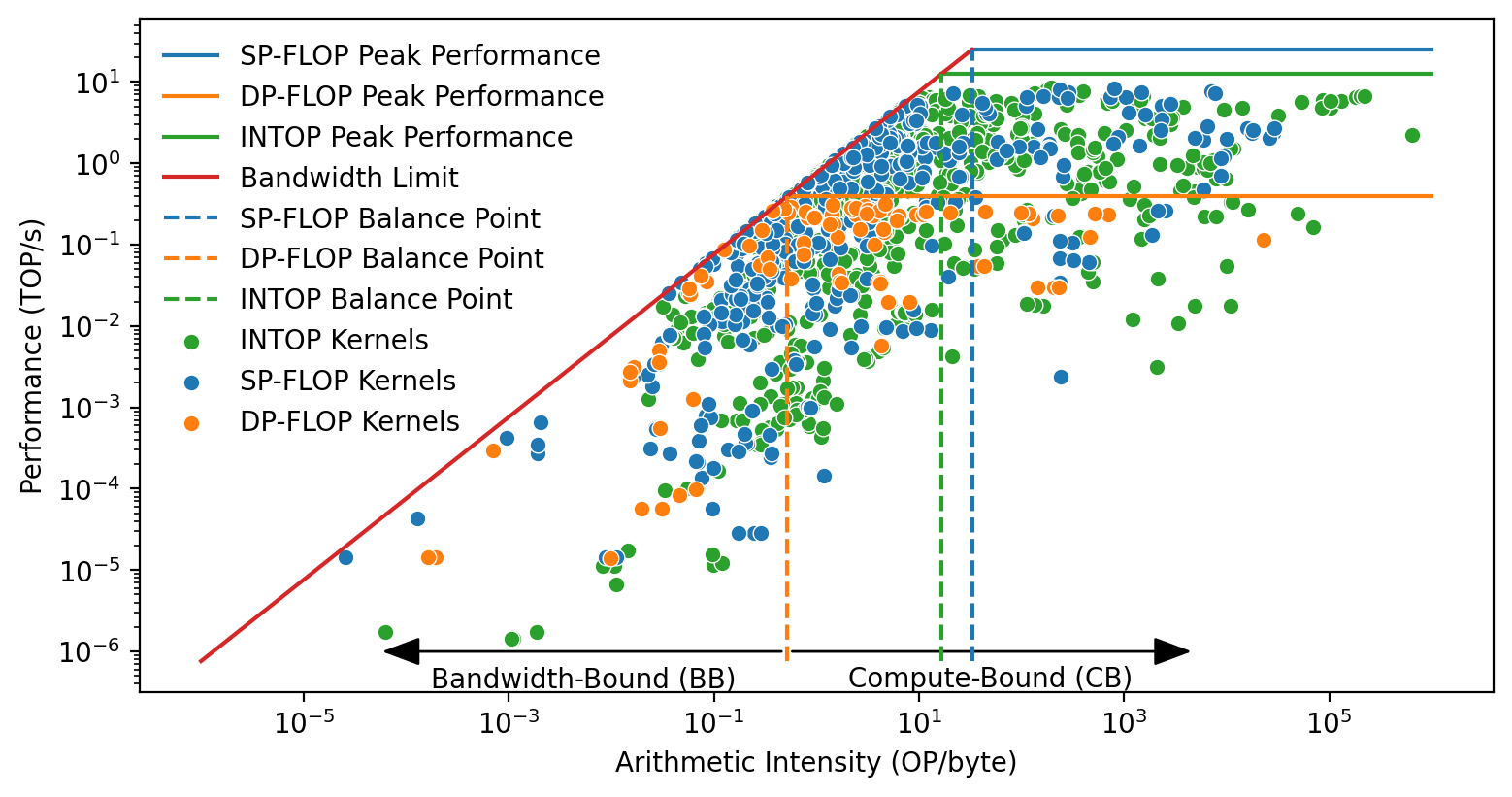}
     \caption{RTX 3080 Roofline Model and profiled HeCBench executable metrics. CB and BB classifications are shown for DP-FLOP balance point.}
     \label{fig:rtx3080Roofline}
 \end{figure}

\autoref{fig:rtx3080Roofline} shows the roofline data of each of the sampled kernels for the target hardware.
The rooflines for SP, DP, and INT ops are shown, along with their respective balance points and the profiled kernel samples as points on the plot.
For all sampled kernels, the theoretical peak performance is usually unmet, as it is often not attainable \cite{rooflineERT1,rooflineERT2} without hardware-specific code modifications to take full advantage of the GPU architecture.
Lastly, we note that the majority of the SP-FLOP and INT samples are BB on this hardware.

\subsection{Dataset Creation \& Pruning}
\label{sec:datasetCreation}

\begin{table*}[ht!]
\centering
\setlength{\tabcolsep}{6pt} 
\renewcommand{\arraystretch}{1.2} 
\resizebox{\linewidth}{!}{%
\begin{tabular}{lcccccccccc}
\toprule
\textbf{Model Name} &
  \textbf{Reasoning} &
  \textbf{\begin{tabular}[c]{@{}c@{}}Input/Output\\ Cost (1M tokens)\end{tabular}} &
  \textbf{RQ1 Acc.} &
  \textbf{RQ1 CoT Acc.} &
  \textbf{RQ2 Acc.} &
  \textbf{RQ2 F1} &
  \textbf{RQ2 MCC} &
  \textbf{RQ3 Acc.} &
  \textbf{RQ3 F1} &
  \textbf{RQ3 MCC} \\
\midrule
\texttt{o3-mini-high} & \checkmark & \$1.1 / \$4.4 & \textbf{100} & \textbf{100} & \textbf{64.12} & \textbf{62.33} & 31.36 & \textbf{63.53} & \textbf{60.91} & \textbf{31.63} \\
\texttt{o1} & \checkmark & \$15 / \$60 & -- & -- & 64.12 & 61.67 & \textbf{32.73} & 61.47 & 58.77 & 26.70 \\
\texttt{o3-mini} & \checkmark & \$1.1 / \$4.4 & \textbf{100} & \textbf{100} & 62.06 & 60.80 & 25.84 & 62.94 & 60.88 & 29.13 \\
\texttt{gpt-4.5-preview} & & \$75 / \$150 & -- & -- & 59.71 & 59.45 & 19.66 & 60.88 & 60.25 & 22.50 \\
\texttt{o1-mini-2024-09-12} & \checkmark & \$1.1 / \$4.4 & \textbf{100} & \textbf{100} & 59.64 & 58.91 & 19.92 & 56.47 & 55.98 & 13.24 \\
\texttt{gemini-2.0-flash-001} & & \$0.1 / \$0.4 & 91.25 & 92.50 & 55.59 & 55.45 & 11.25 & 53.82 & 48.96 & 9.72 \\
\texttt{gpt-4o-2024-11-20} & & \$2.5 / \$10 & 91.25 & 96.25 & 52.06 & 41.04 & 8.20 & 53.24 & 44.17 & 10.93 \\
\texttt{gpt-4o-mini} & & \$0.15 / \$0.6 & 90.00 & \textbf{100} & 50.59 & 50.03 & 1.20 & 52.35 & 50.92 & 5.01 \\
\texttt{gpt-4o-mini-2024-07-18} & & \$0.15 / \$0.6 & 90.00 & \textbf{100} & 50.29 & 49.88 & 0.60 & 52.06 & 50.46 & 4.41 \\
\bottomrule
\end{tabular}%
}
\caption{Evaluation results sorted by RQ1 Accuracy. Cost (as of April 2025), reasoning capabilities, and metrics across RQ1–RQ3 are shown. All values are percentages unless otherwise noted.}
\label{tab:evalResultsRQ1RQ2}
\end{table*}

Due to the imbalance of BB kernels that could bias our reported results, we take an extra pruning step to create a balanced inference dataset.
This pruning step consists of dropping programs with long source code inputs, and dropping excess BB codes.

\noindent\textbf{Source Scraping.}
For the source code that we feed to the LLMs for prediction, we opted for the easiest approach: to concatenate all the source files for each program into one string that is appended to the end of the LLM query prompt. 
Initial attempts to extract only the source code of each kernel proved complicated due to many edge cases related to templating, preprocessor defines, and determining critical execution path that were too challenging to handle.

\noindent\textbf{Token Count Pruning.}
Prompting the LLMs with full source codes leads to token imbalance across prompts.
Given that the LLM context can get lost when prompted with long inputs \cite{needleInHaystackPaper}, LLMs can provide better responses about shorter programs.
To homogenize queried source codes and drop long inputs, we set a cutoff of 8e3 tokens, determined after checking the distribution of tokens and source codes using the \gptFOURoMini tokenizer.
This cutoff allowed us to keep a little more than 50\% of all the profiled kernels, leaving us with 297 CUDA-based and 242 OMP-based sampled HeCBench programs.

\noindent\textbf{Dataset Balancing.}
Because we feed the entire source code of a program when querying an LLM, we risk an unbalanced database if we choose to query with two different kernels from the same program. 
To avoid biasing our evaluation metrics, we only query LLMs about the first kernel from the object dump of each program; thus, each program appears only once in the dataset.

The final balancing step was to force the number of samples for combinations of code \emph{language} (CUDA/OMP) and \emph{class} (BB/CB) to be equal to the smallest set of said combinations.
The smallest combination totaled 85 samples, for a final dataset of 340 samples.
For later fine-tuning in \fineTuningQuestion, we further divide our dataset with an 80/20 training/validation split.
This gave us 68 samples for each language/class \emph{training} combo, and similarly 17 samples for \emph{validation} combos.

\begin{figure}[!ht]
    \centering
    \includegraphics[width=1.0\linewidth]{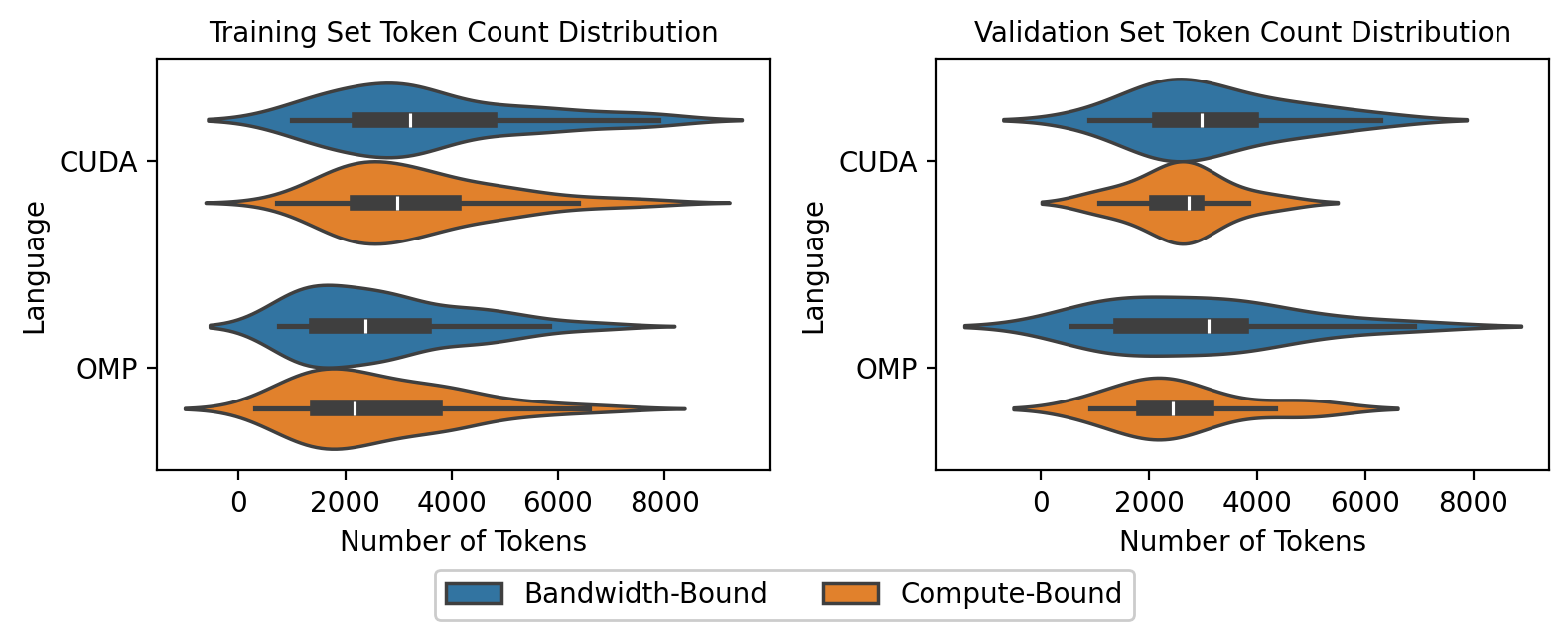}
    \caption{Training/Validation set token count distributions, showing a reasonably balanced dataset.}
    \label{fig:trainTestTokenDistrib}
\end{figure}

\autoref{fig:trainTestTokenDistrib} shows the \gptFOURoMini token distribution of the balanced dataset with box-and-whisker plots.
The OMP codes, are on average, able to use less tokens than the CUDA codes, while the IQR ranges and medians mostly line up. 
Given the small size of the validation set, ensuring balance without dropping too many samples was difficult, so we end up with the CB samples having a slightly tighter range than the BB samples.

\section{Evaluation}

\subsection{Evaluation Metrics}
Because we are performing a binary classification task with LLMs on a balanced dataset, we use the metrics of \textbf{\textit{accuracy}} and \textbf{\textit{F1-score}} (macro variant), as they do not depend on defining a true-positive and true-negative class, which could be either of the CB or BB classes. 
For the same class definition reason, we also use the \textbf{\textit{Matthews Correlation Coefficient}} (MCC) metric.
For the MCC, the metric values range from +1 (for perfect prediction), -1 (for inverse prediction), and 0 (for an average random prediction).
Because all these metric values range from 0 to 1 or -1 to 1, we multiply them by 100 for readability purposes.

\subsection{Sampled LLMs}
\autoref{tab:evalResultsRQ1RQ2} lists the various LLMs we sampled for this work.
The LLMs fall into one of two categories: reasoning and non-reasoning LLMs. 
We mainly sample OpenAI LLMs because of their popularity in other work, as well as their ease of access through Microsoft Azure's OpenAI LLM services \cite{microsoftAzureOpenAI}.
As an additional comparison point, we also sample Google's Gemini Flash 2.0 \cite{blogIntroducingGeminiFlash2.0}. 
Our objective is to see if LLMs can even be good roofline performance classifiers, thus we only sample a handful of LLMs that are representative of the current SoTA.

\noindent\textbf{LLM Hyperparameters.}
LLMs come with various sampling hyperparameters that can affect their generated output.
The main hyperparmeters are: \textbf{\texttt{temperature}} and \textbf{\texttt{top\_p}}, which control the variety of the model responses, and limits the model's output token choices, respectively.
We conducted a Chi-Squared test on the LLMs listed in \autoref{tab:evalResultsRQ1RQ2} and found that a change in these two hyperparameters did not have any statistically significant impact on the predicted outcomes of the LLMs \cite{modernLLMsAreTempInvariant}.
For all further tests, we set the \texttt{temperature} to 0.1 and \texttt{top\_p} to 0.2, which essentially guarantees less diverse and consistent model responses.
The reasoning models in \autoref{tab:evalResultsRQ1RQ2} do not allow for changes in the sampling hyperparameters, therefore, we query them with their default settings.

\subsection{Experimental Design}
We conducted one experiment for each research question from \autoref{sec:RQs}; details are provided below.

\noindent\textbf{\classificationQuestion: Baseline Roofline Calculations.}
To understand whether LLMs were capable of roofline model calculations for \textit{arbitrary} hardware, we first tried a simple experiment of querying them with random rooflines accompanied by random AI values for classification. 
Given a roofline, the LLMs are expected to derive a balance point that separates BB from CB codes on the AI axis, allowing them to easily categorize requested AI values.
We randomly generated 240 different rooflines, where we would randomly select a BB and a CB AI for classification.

\autoref{fig:cotPrompt} shows the provided prompt, where the highlighted regions show values that were changed between prompts according to the random roofline used. 
The prompt was designed to show 2, 4, or 8 examples, with (and without) chain-of-thought (CoT) \cite{cotPrompting} thought text, demonstrating how the model should structure its output and perform reasoning to arrive at a classification.
To avoid erratic responses and allow for easy automation of response correctness checking, we purposely include at least \emph{two} question and response examples in the prompts. 

\begin{figure}[!ht]
    \centering
    \begin{tcolorbox}[
        colframe=OliveGreen!70!black,     
        colback=white,                    
        coltitle=white,                   
        colbacktitle=OliveGreen!80!black,
        title=Chain-of-Thought (CoT) Prompt Example,
        fonttitle=\bfseries,
        enhanced,
        sharp corners=south,
        boxrule=0.4mm,
        width=1.0\linewidth,
        before upper={\parindent15pt},    
        fontupper=\small,
        fontlower=\tiny,
        left=1pt,
        right=1pt,
        top=1pt,
        bottom=0pt,
        arc=2mm,
        drop shadow=black!50!white,
    ]

\hl{CoT example 1 (shown below):}

\noindent\textbf{Question:} Given a GPU having a global memory with a max bandwidth of \hl{45.9} GB/s and a peak performance of \hl{52.22} GFLOP/s, if a program executed with an Arithmetic Intensity of \hl{0.6} FLOP/Byte and a performance of \hl{19.4} GFLOP/s, does the roofline model consider the program as compute-bound or bandwidth-bound? 

\noindent\textbf{Thought:} The max bandwidth is \hl{45.9} GB/s, and peak performance is \hl{52.22} GFLOP/s. The balance point is at \hl{52.22 / 45.9}= \hl{1.14} FLOP/Byte. The program's Arithmetic Intensity is \hl{0.6} FLOP/Byte. 
Because \hl{0.6 < 1.14}, it is \hl{before} the balance point, putting the program in the \hl{bandwidth-bound} region. The roofline model would consider the program as \hl{bandwidth-bound}.

\noindent\textbf{Answer:} \hl{Bandwidth}

\vspace{0.5em}
\hl{CoT examples 2-8 [redacted]}

\vspace{0.5em}
\noindent\textbf{Question:} Given a GPU having a global memory with a max bandwidth of \hl{99.9} GB/s and a peak performance of \hl{73.45} GFLOP/s, if a program executed with an Arithmetic Intensity of \hl{1.55} FLOP/Byte and a performance of \hl{32.8} GFLOP/s, does the roofline model consider the program as compute-bound or bandwidth-bound?

    \end{tcolorbox}
    \caption{Prompt for \classificationQuestion. Highlighted text indicates values changed between invocations.}
    \label{fig:cotPrompt}
\end{figure}

\noindent\textbf{\zeroShotQuestion: Zero-Shot.}
For \zeroShotQuestion, we increase the roofline classification task difficulty from \classificationQuestion by querying the LLMs without profiling information, instead we provide hardware information and source code to be classified, with minimal instructions and pseudo-code classification examples.

\begin{figure}[!ht]
    \centering
    \begin{tcolorbox}[
        colframe=OliveGreen!70!black,
        colback=white,
        coltitle=white,
        colbacktitle=OliveGreen!80!black,
        title=System Prompt for Classifying GPU Kernels (RQ1–RQ3),
        fonttitle=\bfseries,
        enhanced,
        sharp corners=south,
        width=1.0\linewidth,
        fontupper=\small,
        fontlower=\tiny,
        left=1pt,
        right=1pt,
        top=1pt,
        bottom=0pt,
        arc=2mm,
        boxrule=0.4mm,
        drop shadow southeast,
    ]

You are a \textbf{GPU performance analysis expert} that classifies kernels into Arithmetic Intensity Roofline model categories based on their source code characteristics. Your task is to provide one of the following performance boundedness classifications: \textbf{Compute} or \textbf{Bandwidth}.

A kernel is considered \textbf{Compute bound} if its performance is primarily limited by the number of operations it performs, and \textbf{Bandwidth bound} if its performance is primarily limited by the rate at which data can be moved between memory and processing units.

\vspace{0.5em}
Provide \textbf{only one word} as your response, chosen from the set: \texttt{['Compute', 'Bandwidth']}.

\vspace{1em}
\textbf{Examples:}

\vspace{0.5em}
\textbf{Example 1:} \hl{[changed to real CUDA/OMP code for RQ2]}
\begin{tcolorbox}[colback=gray!10, boxrule=0pt, top=0pt, bottom=0pt, left=3pt, right=3pt, enhanced]
\texttt{Kernel Source Code (simplified):}\\
\texttt{for i = 0 to 1000000 \{}\\
\indent\texttt{~~a[i] = a[i] + b[i];}\\
\texttt{\}}
\end{tcolorbox}
Response: \textbf{Compute}

\vspace{0.5em}
\textbf{Example 2:} \hl{[changed to real CUDA/OMP code for RQ2]}
\begin{tcolorbox}[colback=gray!10, boxrule=0pt, top=0pt, bottom=0pt, left=3pt, right=3pt, enhanced]
\texttt{Kernel Source Code (simplified):}\\
\texttt{for i = 0 to 10 \{}\\
    \indent\texttt{~~load\_data(large\_array);}\\
    \indent\texttt{~~process\_data(large\_array);}\\
    \indent\texttt{~~store\_data(large\_array);}\\
\texttt{\}}
\end{tcolorbox}
Response: \textbf{Bandwidth}

\vspace{1em}
Now, analyze the following source codes for the requested kernel of the specified hardware.

Classify the \hl{[language]} kernel called \hl{[kernel name]} as \textbf{Bandwidth} or \textbf{Compute} bound. The system it will execute on is a \hl{[GPU model]} with:
\begin{itemize}
    \item peak single-precision performance of \hl{[X]} GFLOP/s
    \item peak double-precision performance of \hl{[X]} GFLOP/s
    \item peak integer performance of \hl{[X]} GINTOP/s
    \item max bandwidth of \hl{[X]} GB/s
\end{itemize}

The block and grid sizes of the invoked kernel are \hl{(X,Y,Z)} and \hl{(X,Y,Z)}, respectively.  
The executable running this kernel is launched with the following command-line arguments: \hl{[\texttt{arg1 arg2 arg3}]}.

Below is the source code of the requested \hl{[language]} kernel:

\vspace{0.3em}
\hl{[concatenated source code files]}

    \end{tcolorbox}
    \caption{System prompt to query LLMs. Highlighted text changes based on the queried source code and hardware.}
    \label{fig:sysPrompt}
\end{figure}

\autoref{fig:sysPrompt} shows the prompt we used to query the LLMs.
We found the pseudo-code examples in the prompt provided enough context for smaller LLMs to produce only the desired output singleton tokens of \textit{Compute} or \textit{Bandwidth}, corresponding to CB and BB classifications, respectively.
The highlighted portions of: \textit{language} (CUDA or OMP), GPU hardware information, and source code change between invocations.
We query all 340 profiled programs from the dataset defined in \autoref{sec:datasetCreation}.

\noindent\textbf{\fewShotQuestion: Few-Shot.}
Similar to \zeroShotQuestion, this experiment uses the same prompt but replaces the pseudo-code examples with actual code examples which can be found on our repository \cite{ourGithub}.
These real examples are not part of the dataset from \autoref{sec:datasetCreation}.
In testing, we noticed that supplying more than the two classification examples led to excessively bloated input prompts that easily lost contextual information, thus we only query with two examples, effectively making our few-shot experiments \textit{two-shot}.  
For \fewShotQuestion, we again query all 340 profiled programs from \autoref{sec:datasetCreation}, where we make sure to only supply examples in the language (CUDA/OMP) of the queried source code.

\noindent\textbf{\fineTuningQuestion: Fine-tuned.}
We wanted to see how well a model performs when fine-tuned with the roofline dataset we created.
We used the Microsoft Azure OpenAI services \cite{microsoftAzureOpenAI} to train \gptFOURoMini-2024-07-18 with the 80/20 train/test split dataset from \autoref{sec:datasetCreation}.
It is typical to train LLMs on datasets with thousands of samples, unfortunately it is difficult to gather more samples while maintaining a balanced dataset, so we experimented with what we had. 
We leave it as future work to augment the dataset for more holistic training.

\subsection{\classificationQuestion: Roofline Calculation Results}
Columns 4 and 5 of \autoref{tab:evalResultsRQ1RQ2} show the best accuracy metrics from the 2, 4, and 8-shot prompting of the LLMs as in \autoref{fig:cotPrompt}. 
All the LLMs achieve a high accuracy between 90-100\%, where the COT prompting notably helps the non-reasoning LLMs.
Across all the few-shot CoT prompts, the reasoning LLMs score 100\% accuracy.
We omit the metrics of F1 and MCC from \autoref{tab:evalResultsRQ1RQ2} because they exhibit the same trends as the accuracy.
The results for \gptOone and \gptFOURFIVE are excluded because their smaller counterparts already perform so well.

\begin{center}
\begin{tcolorbox}[
    colframe=OliveGreen!80!black,
    colback=SpringGreen!20!white,
    width=\columnwidth,
    boxrule=0.3mm,
        left=1pt,
        right=1pt,
        top=1pt,
        bottom=0pt,
        arc=2mm,
    enhanced,
    drop shadow southeast,
    sharp corners=south,
    nobeforeafter,
    fontupper=\small,
    title={\textbf{Answer to \classificationQuestion}},
]

Existing SoTA LLMs are capable of making sense of roofline problems and categorizing a program's arithmetic intensity when given all the GPU roofline specifications and program profiling results.

\end{tcolorbox}
\end{center}

\subsection{\zeroShotQuestion: Zero-Shot Results}
Columns 6, 7, and 8 of \autoref{tab:evalResultsRQ1RQ2} show the results for the zero-shot experiments.
The best models in terms of \textit{accuracy} are the \gptOthreeMiniHigh and \gptOone, both correctly predicting the class of 64\% of the source code samples.
Due to only having accuracy differences of on average 5\% when solely predicting CUDA or OMP codes, we omit the metrics for predictions across language, instead we present joint metrics.
Note that there is a clear performance difference between the reasoning and non-reasoning models, where the non-reasoning LLMs make predictions akin to a random predictor, given the accuracy values near 50\% and MCC values close to 0.
The reasoning LLMs are able to achieve about a 10\% greater accuracy over the non-reasoning LLMs, indicating that the extra reasoning steps of the models are contributing to better predictions.

\begin{center}
\begin{tcolorbox}[
    colframe=OliveGreen!80!black,
    colback=SpringGreen!20!white,
    width=\columnwidth,
    boxrule=0.3mm,
        left=1pt,
        right=1pt,
        top=1pt,
        bottom=0pt,
        arc=2mm,
    enhanced,
    drop shadow southeast,
    sharp corners=south,
    nobeforeafter,
    fontupper=\small,
    title={\textbf{Answer to \zeroShotQuestion}},
]
LLMs \textit{can} predict a CUDA/OMP program's arithmetic intensity classification, especially when using reasoning-based models.
However, there is still room for improvement, with the best model achieving a maximum accuracy of 64\%. 
\end{tcolorbox}
\end{center}

\subsection{\fewShotQuestion: Few-Shot Results}
Similar to \zeroShotQuestion, columns 9, 10, and 11 of \autoref{tab:evalResultsRQ1RQ2} show the results of the few-shot experiments.
We can notice that again \gptOthreeMiniHigh and \gptOone end up on top across all three metrics.
For the reasoning models, there is not much of a difference in metrics between the few-shot and zero-shot experiments.
However, the small (\textit{mini}) non-reasoning LLMs perform marginally better (2\% accuracy improvement) given the real examples in the prompt.

\begin{center}
\begin{tcolorbox}[
    colframe=OliveGreen!80!black,
    colback=SpringGreen!20!white,
    width=\columnwidth,
    boxrule=0.3mm,
        left=1pt,
        right=1pt,
        top=1pt,
        bottom=0pt,
        arc=2mm,
    enhanced,
    drop shadow southeast,
    sharp corners=south,
    nobeforeafter,
    fontupper=\small,
    title={\textbf{Answer to \fewShotQuestion}},
]
Few-shot examples yield marginal improvements on the metrics of non-reasoning models, while hardly impacting reasoning-based LLMs.
Our recommendation would be to save money on input token costs by prompting in zero-shot style with reasoning models.
\end{tcolorbox}
\end{center}

\subsection{\fineTuningQuestion: Fine-Tuning Results}
For model fine-tuning, the outcome was poor.
We fine-tuned \gptFOURoMini for two epochs, where at the end of both epochs, the model had devolved and would always predict either CB or BB for the whole validation set.
The same behavior would occur when fine-tuned solely on CUDA or OMP programs.
The training prompts used were the zero-shot prompts from \zeroShotQuestion, as we wanted to see if we could achieve an accuracy above 50\%. 
We strongly suspect the fine-tuned model did not have enough training samples to guarantee generalization, which caused it to produce incorrect predictions.

\begin{center}
\begin{tcolorbox}[
    colframe=OliveGreen!80!black,
    colback=SpringGreen!20!white,
    width=\columnwidth,
    boxrule=0.3mm,
        left=1pt,
        right=1pt,
        top=1pt,
        bottom=0pt,
        arc=2mm,
    enhanced,
    drop shadow southeast,
    sharp corners=south,
    nobeforeafter,
    fontupper=\small,
    title={\textbf{Answer to \fineTuningQuestion}},
]
Our current results indicate fine-tuning completely biases LLMs to always answer either CB or BB. 
We speculate that a larger training dataset is necessary to guarantee a more generalizable fine-tuned LLM.
\end{tcolorbox}
\end{center}

\section{Conclusion \& Future Work}

In this paper, we explore the applicability of LLMs for classifying the arithmetic intensity (AI) of CUDA-based and OpenMP-based GPU source codes.
We discover that LLMs indeed can predict whether a program is \CB (CB) or \BB (BB) without the need of execution/profiling.
A distinction between reasoning-based models and non-reasoning models arises, where reasoning LLMs significantly outperform non-reasoning models.
We show that with some simple prompting techniques, we are able to get LLMs to predict with at best a 64\% accuracy. 

\noindent \textbf{Expanding Dataset.} 
Given different GPU hardware, the arithmetic intensity of a program may change from CB to BB. 
Our experimentation was performed with a single GPU.
To have more generalizable results, it would be best to re-profile all our GPU programs on varying hardware to see how LLM prediction accuracy changes.
This would expand our dataset size to also allow for more robust fine-tuning using thousands of samples, rather than hundreds.
We would be able to make more conclusive statements about whether fine-tuning an LLM to this problem is worth the effort.

\noindent \textbf{Improve Prediction Accuracy.}
Our evaluation of LLMs using prompting techniques demonstrated promise. 
However, more recent question-decomposition \cite{llmQuestionDecomposition}, successive-prompting \cite{successivePrompting}, and least-to-most \cite{leastToMostPrompting} prompting techniques have shown effectiveness in breaking down and solving complex tasks.
In an effort to improve roofline classification metrics, these techniques warrant further investigation.

\newpage
\begin{acks}
This research was supported by Code Metal Inc. and the Pazy foundation. This work was performed under the auspices of the U.S. Department of Energy by Lawrence Livermore National Laboratory under Contract DE-AC52-07NA27344.
\end{acks}

\bibliographystyle{ACM-Reference-Format}
\bibliography{references}

\appendix

\end{document}